\documentclass[12pt]{article}
\usepackage{epsfig}

\newcommand{\List}[2]{\begin{list}{$\bullet$}{
\setlength{\labelwidth}{#2}
\setlength{\leftmargin}{\labelwidth}\addtolength{\leftmargin}{#1} 
}}
\def\Item[#1]{\item[#1\hfill]}

\newif\ifvecarrow
\newif\ifvecboldmath
\newcommand{\beq}{
  \begin{eqnarray}}
\newcommand{\eeq}{\end{eqnarray}
   }
\newcommand{\bdis}{
  \begin{eqnarray*}}
\newcommand{\edis}{\end{eqnarray*}
   }
\newcommand{\eq}[1]{eq.(\protect\ref{#1})}

\newdimen\dmove \newdimen\vmove
   \newcount\moven \newcount\movenh  \newcount\movenv
\let \vecarrow=\vec
\newcommand{\vecboldmath}[1]{\mathchoice
   {\mbox{\boldmath $#1$}}
   {\mbox{\boldmath $#1$}}
   {\mbox{\footnotesize\boldmath $#1$}}
   {\mbox{\footnotesize\boldmath $#1$}}}
\renewcommand{\vec}[1]{{\ifvecarrow \vecarrow{#1} 
             \else \ifvecboldmath \vecboldmath{#1}
                    \else {\bf {#1}} \fi \fi}}
\newcommand{\vect}[1]{
  \ifvecarrow 
    \vecarrow{#1}
  \else  
    \ifvecboldmath 
       \vecboldmath{#1} 
    \else
       \mathchoice{\global\dmove=0.06ex\global\setbox0=\hbox{{$ #1$}}\vecto}
                  {\global\dmove=0.06ex\global\setbox0=\hbox{{$ #1$}}\vecto}
        {\global\dmove=0.042ex\global\setbox0=\hbox{{$\scriptstyle #1$}}\vecto}
        {\global\dmove=0.03ex\global\setbox0=\hbox{{$\scriptscriptstyle #1$}}\vecto}
    \fi  
  \fi}
\newcommand{\vecto}{\moven=5\movenh=0\divide\dmove by\moven\loop%
 {\vmove=\dmove \movenv=\moven\advance\movenv by -1\divide\movenv by 2%
  \multiply\vmove by -\movenv \movenv=0 \loop%
      \raise\vmove\copy0\kern-\wd0%
   \advance\movenv by 1\ifnum\movenv<\moven \advance\vmove by \dmove \repeat}%
   \advance\movenh by 1\ifnum\movenh<\moven\kern\dmove\repeat
      \kern\wd0} 

\newcommand{\abs}[1]{\left\vert #1 \right\vert}

\catcode`\@=11
\def\bracket#1{\@ifnextchar[{\bracketXXX{#1}}{\bracketXX{#1}}}
\def\bracketXXX#1[#2]#3{%
\left\langle{#1}\kern0.2em\vrule\kern0.2em{#2}\kern0.2em\vrule\kern0.2em{#3}%
\right\rangle}
\def\bracketXX#1#2{%
\left\langle{#1}\kern0.2em\vrule\kern0.2em{#2}\right\rangle}

\catcode`\@=12

\newcommand{\scp}{{\mathchoice{\cdot}{\cdot}{\cdot}{\cdot}}}

\newcommand{\avr}[1]{\left< #1 \right>}
\catcode`\@=11
\def\lsim{\mathrel{\mathpalette\@versim<}}
\def\gsim{\mathrel{\mathpalette\@versim>}}
\def\@versim#1#2{\vcenter{\offinterlineskip
	\ialign{$\m@th#1\hfil##\hfil$\crcr#2\crcr\sim\crcr } }}
\catcode`\@=12

\newcommand{\lemma}[2]{\vskip 0.5ex \noindent\hbox to 5mm{}{\bf [#1]}
\hbox to 5mm{}\parbox[t]{140mm}{{\it #2}}\vskip 0.5ex\noindent}
\newcommand{\M}{{\scriptstyle{M}}}

\begin{document}

\baselineskip 3ex

\begin{flushleft}
DESY 99-006\hfill ISSN 0418-9833\\
January 1999
\end{flushleft}
\begin{center}
{\Large An Algorithm for Calculating the Spin Tune in Circular Accelerators}
\end{center}
\vskip 10mm
\begin{center}
           {\Large Kaoru Yokoya}\footnote{
 On leave of absence from the National Accelerator Research Organization, Japan
(KEK).}\\
  Deutsches Electronen Synchrotron, DESY, Hamburg, Germany
\end{center}
\vskip 20mm
\begin{center}
   {\bf Abstract}
\end{center}

A new algorithm for calculating the spin tune and the $\vect{n}$-axis
for circular accelerators is presented. The method resembles
the one employed in the existing program code SODOM in that
one-turn numerical spin maps at equally spaced orbit angle variables
are used but it is more efficient than the latter. Furthermore, it is applicable
at large openning angles of the $\vect{n}$-axis, whereas the existing SODOM only
converges for small angles.

\newpage
\section{Introduction}
  The spin tune, the spin precession frequency divided by the orbit
revolution frequency, is an important parameter in the description of
spin motion in circular accelerators. When a particle is on the
closed orbit, the definition of the spin tune is obvious; 
it is the spin precession
angle over one turn divided by $2\pi$. However, when orbit
oscillations are involved, the definition of the spin tune
becomes more complicated. One needs the concept of the $\vect{n}$-axis 
which was first introduced by Derbenev and Kondratenko\cite{DK73}
for radiative polarization phenomena in electron storage rings.

   We assume that we have complete knowledge about the orbit motion,
i.e. that we know the action and angle variables, 
$\vect{J}=(J_x,J_y,J_z)$ and $\vect{\phi}=(\phi_x,\phi_y,\phi_z)$,
corresponding to the three degrees of freedom of the orbit motion,
which can be in general nonlinear.

   A particle with initial coordinates $(\vect{J},\vect{\phi})$
at a machine azimuth  $\theta$  executes orbit oscillations and comes to
$(\vect{J},\vect{\phi}+\vect{\mu})$ after one turn 
($\theta \rightarrow \theta+2\pi$), where
$\vect{\mu}=(\mu_x,\mu_y,\mu_z)$ is the orbit tune, $\vect{\nu}$, times $2\pi$.
The spin motion over one turn can in general be expressed by a 3$\times$3
rotation matrix $R(\vect{J},\vect{\phi},\theta)$. Obviously, it is
a periodic function of $\theta$ and $\vect{\phi}$ with period $2\pi$.
On the next turn the rotation is expressed by 
$R(\vect{J},\vect{\phi}+\vect{\mu},\theta+2\pi)$
= $R(\vect{J},\vect{\phi}+\vect{\mu},\theta)$
 which differs from 
$R(\vect{J},\vect{\phi},\theta)$ unless the orbit tunes are integers. 

A particle on the closed orbit sees the same rotation $R_0(\theta)$ 
for every turn.
$R_0(\theta)$ has eigenvalues $1$ and $e^{\pm i\mu_{s0}}$ and 
the spin tune  $\nu_{s0}$ is $\mu_{s0}/2\pi$. One can show that $\mu_{s0}$ is independent
of $\theta$.
The eigenvector belonging to the eigenvalue
$1$ is denoted by $\vect{n}_0$, i.e., $R_0\vect{n}_0=\vect{n}_0$. 
It depends only on $\theta$.
 A spin parallel to $\vect{n}_0(\theta)$ remains unchanged
after one turn, and all other spins attached to closed orbit trajectories 
precess by the angle $\mu_{s0}$
around $\vect{n}_0$ during one turn.

The vector $\vect{n}$ is a generalization of $\vect{n}_0$
for particles off the closed orbit. It is 
a function of $(\vect{J},\vect{\phi},\theta)$ periodic in
$\vect{\phi}$ and $\theta$ and satisfies
\beq
   R(\vect{J},\vect{\phi},\theta) \vect{n}(\vect{J},\vect{\phi},\theta)
       = \vect{n}(\vect{J},\vect{\phi}+\vect{\mu},\theta).
\eeq
When $\vect{J}=0$, $\vect{n}$ reduces to $\vect{n}_0$.
To define the spin tune
for nonzero $\vect{J}$, we need two more vectors $\vect{u}_1$ and $\vect{u}_2$
which form an orthonormal basis together with $\vect{n}$. They
are functions of $(\vect{J},\vect{\phi},\theta)$ and periodic in
$\vect{\phi}$ and $\theta$ like $\vect{n}$. The spin tune is
defined as the precession angle in the frame 
$(\vect{u}_1,\vect{u}_2,\vect{n})$ divided by $2\pi$.

   The concept of the vector $\vect{n}$ has been playing an
important role in the description and calculation of radiative
polarization in electron/positron storage rings since\cite{DK73}.
Recently, it has also turned out to be useful for proton rings\cite{DESY}.

   To calculate the vector $\vect{n}$ several
algorithms have been invented. S.~Mane\cite{SMILE} developed a computer
code SMILE using a perturbation expansion with respect to the
orbit action variable. The present author suggested a perturbation
algorithm using Lie algebra\cite{KYLie} and 
Eidelmann and Yakimenko\cite{SPINLIE}
coded a program SPINLIE with (low order) orbit nonlinearity.
Balandin, Golubeva and Barber\cite{BG} also wrote a Lie Algebra code.

The present author considered another method which does not employ
a perturbation expansion and wrote a program SODOM\cite{SODOM}.
Heinemann and Hoffstaetter\cite{SPRINT} use tracking and 
`stroboscopic averaging' in the code SPRINT.
The programs SODOM, SPRINT and \cite{BG} additionally compute the spin tune.

The new method which we are going to describe is based on SODOM.

   We shall briefly summarize the SODOM algorithm in the next
section and describe the new method in Sec.3.

\section{The SODOM Algorithm}
  Let us first briefly summarize the algorithm employed in SODOM.
(See Sec.3 of \cite{SODOM}.)  Denote the one-turn SU2 spin transport map
starting at a fixed prechosen azimuth $\theta_0$
for particles with initial orbital phase $\vect{\phi}$
by $M(\vect{\phi})$ and the spinor representing the $\vect{n}$-axis at 
$\theta_0$
by $\psi(\vect{\phi})$. (Here, we simply write $\psi(\vect{\phi})$
instead of $\psi_{+}(\vect{\phi})$ \cite{SODOM}. We also 
omit the arguments $\vect{J}$ and $\theta_0$ since we shall deal with
one set of $\vect{J}$ and consider the one-turn map from the 
origin $\theta_0$ only.)
The fact that $\vect{n}$ is `invariant'
means
\beq
     M(\vect{\phi})\psi(\vect{\phi})=e^{-iv(\vect{\phi})/2}
         \psi(\vect{\phi}+\vect{\mu}),
\eeq
where $v(\vect{\phi})$ is a real periodic function. Once a solution
($\psi(\vect{\phi})$,$v(\vect{\phi})$) is obtained,
we solve the equation
\beq
    v(\vect{\phi}) + u(\vect{\phi}+\vect{\mu}) - u(\vect{\phi}) = \mu_s
\eeq
and  define 
\beq
    \Psi(\vect{\phi})\equiv e^{iu(\vect{\phi})/2}\psi(\vect{\phi})
\eeq
Then, $\Psi(\vect{\phi})$ satisfies
\beq
     M(\vect{\phi})\Psi(\vect{\phi})=e^{-i\mu_s/2}
         \Psi(\vect{\phi}+\vect{\mu}),
         \label{Psieq}
\eeq
where $\mu_s$ is the spin tune times $2\pi$.
  The $\vect{u}_{1,2}$ axes are represented by a spinor  
\beq
   \Psi_{\varphi}\equiv \frac{1}{\sqrt{2}} \left[ 
      e^{-i\varphi/2}\Psi+e^{i\varphi/2} \widehat{\Psi}^{*} \right] 
          \label{Psivarphi}
\eeq
where we define the operation $\widehat{\ }$ as
\beq
     \widehat{\Psi} \equiv i\sigma_2 \Psi^{*},
\eeq
which was denoted by $\Psi_{-}$ in \cite{SODOM}.
Note that $\widehat{\widehat{\Psi}}=-\Psi$ and 
$\widehat{\Psi}^\dagger\Psi=0$.
The three spinors, $\Psi_0$, $\Psi_{\pi/2}$, $\Psi$, represent the three
vectors $\vect{u}_1$, $\vect{u}_2$, $\vect{n}$. The phase of $\Psi$ is
irrelevant for defining $\vect{n}$ but it is important for $\vect{u}_1$
and $\vect{u}_2$.

The original SODOM algorithm parametrizes $\psi$ as
\beq
   \psi=\frac{1}{\sqrt{1+\abs{\zeta(\vect{\phi})}^2}}
       \left( \begin{array}{c}
                    1  \\  \zeta(\vect{\phi})  \end{array} \right).
       \label{psiwithzeta}
\eeq
The SU2 matrix $M(\vect{\phi})$ can be parametrized by two complex
functions $f(\vect{\phi})$ and $g(\vect{\phi})$ as
\beq
    M(\vect{\phi}) = 
    \pmatrix{ -ig(\vect{\phi}) & -if^{*}(\vect{\phi})  \cr
              -if(\vect{\phi}) & ig^{*}(\vect{\phi})  \cr}
\eeq
Then, one gets an equation for $\zeta$:
\beq
    g^{*}(\vect{\phi}) \zeta(\vect{\phi})
        +g(\vect{\phi}) \zeta(\vect{\phi}+\vect{\mu})
     = f(\vect{\phi}) - f^{*}(\vect{\phi})\zeta(\vect{\phi})
                      \zeta(\vect{\phi}+\vect{\mu}).
       \label{zetaeq}
\eeq
By expanding $f(\vect{\phi})$, $g(\vect{\phi})$, and
$\zeta(\vect{\phi})$ into Fourier series like
$\sum f_{\vect{m}} e^{i\vect{m}\scp\vect{\phi}}$, we get a nonlinear
equation for $\zeta_{\vect{m}}$.

A key component of SODOM is the calculation of the Fourier coefficients 
$f_{\vect{m}}$
and $g_{\vect{m}}$ from the tracking data  over one turn 
for several particles having the same $\vect{J}$ but
equally-spaced $\vect{\phi}$ ($0\leq\phi<2\pi$).
 
  The parametrization (\ref{psiwithzeta})
is good only when $\zeta(\vect{\phi})$ is small. Because of its 
up-down asymmetric form, many more Fourier terms are needed than
required by the physics, when $\zeta(\vect{\phi})$ is large.\footnote{
For example, $\Psi=(\cos\phi,\sin\phi)$ is a mild function but
leads to $\zeta=\tan\phi$ which is hard to Fourier-expand.}
Also, the iterative method of solving the nonlinear equation easily fails
when $\zeta$ is large.

\section{The Matrix Eigenvalue Method}
  The new algorithm is much simpler and involves solving \eq{Psieq}
directly rather than \eq{zetaeq}. 
By expanding $M(\vect{\phi})$ (actually the functions
$f(\vect{\phi})$ and $g(\vect{\phi})$) and $\Psi(\vect{\phi})$ into 
Fourier series as
\beq
    M(\vect{\phi}) = \sum_{\vect{m}} M_{\vect{m}}
            e^{i\vect{m}\scp\vect{\phi}}, \qquad
    \Psi(\vect{\phi}) = \sum_{\vect{m}} \Psi_{\vect{m}}
            e^{i\vect{m}\scp\vect{\phi}}
       \label{FourierEq}
\eeq
\eq{Psieq} can be written as
\beq
   e^{-i\vect{m}\scp\vect{\mu}} \sum_{\vect{m}'}
           M_{\vect{m}-\vect{m}'} \Psi_{\vect{m}'}
      = e^{-i\mu_s/2} \Psi_{\vect{m}}.
            \label{mateq}
\eeq
  This is simply a matrix eigenvalue equation.
Thus, the spin tune comes out as an eigenvalue.

   However, obviously, \eq{mateq} has many eigenvalues.
Which one gives the spin tune?  
What do the other eiganvalues and eigenvectors mean?
In order to answer these questions
  let us return to \eq{Psieq} and examine it as an eigenvalue system
\beq
     M(\vect{\phi})\Psi(\vect{\phi})=\lambda
         \Psi(\vect{\phi}+\vect{\mu})
         \label{Psieqgeneral}
\eeq
Note that this is not a simple 2$\times$2 algebraic equation
because of the $\vect{\phi}+\vect{\mu}$.

    Before going further we have to think about subtle problems associated
with the `2-to-1' correspondence between SU2 and SO3.
Note that we use 2-component spinors and SU2 matrices instead of 3-vectors
and SO3 matrices to achieve computational speed and to minimize storage 
but not because the particles have spin $\hbar/2$.  
The classical spin motion can be
completely described by 3-vectors and SO3 matrices. 

   What does the periodicity of a spinor with respect to $\vect{\phi}$ mean?  
The physical object is the 3-vector $\Psi^\dagger\vect{\sigma}\Psi=\vect{n}$
rather than the spinor $\Psi$. 
In this sense a complex phase factor in $\Psi$ is irrelevant.
However, a complex phase factor is still relevant when one constructs 
the $\vect{u}_1$ and $\vect{u}_2$ axes from $\Psi$ via $\Psi_\varphi$.

  On the other hand, a sign change of $\Psi$ does not cause a change
of $\vect{n}=\Psi^\dagger\vect{\sigma}\Psi$ nor a change of $\vect{u}_1$
and $\vect{u}_2$ defined by $\Psi^\dagger_\varphi\vect{\sigma}\Psi_\varphi$.

  Thus, as the periodicity condition for $\Psi$ with respect to $\phi_j$
(one of the orbit angle variables),
 we have to allow both $\Psi(\phi_j+2\pi)=\Psi(\phi_j)$ and
$\Psi(\phi_j+2\pi)=-\Psi(\phi_j)$. 
Then with 3 degrees of freedom for orbit motion, we have 8 types of solutions
$\Psi(\vect{\phi})$ differing by their sign change behaviour
under the transformation
$\phi_j\rightarrow\phi_j+2\pi$.  In Fourier expansion language, this 
means that $\Psi$ can be expanded as
\beq
     \Psi(\vect{\phi})=e^{i\vect{m}^0\scp\vect{\phi}/2}
           \sum_{\vect{m}} \Psi_{\vect{m}} e^{i\vect{m}\scp\vect{\phi}}
        \label{shiftedFourier}
\eeq
where $\vect{m}^0$ is a set of three integers each of which is either 0 or 1.

  We now define the scalar product of two arbitrary spinors 
$\Psi_1$ and $\Psi_2$ by
\beq
      (\Psi_1,\Psi_2) \equiv \frac{1}{(4\pi)^3}\int_0^{4\pi}
                   \Psi_1^{\dagger}(\vect{\phi})
           \Psi_2(\vect{\phi})  d\vect{\phi}
       = \delta_{\vect{m}^0_1,\vect{m}^0_2}
        \sum_{\vect{m}} \Psi_{1,\vect{m}}^{\dagger} \Psi_{2,\vect{m}}
        \label{defscalarprod}
\eeq
Obviously, solutions of different types in \eq{shiftedFourier} are always orthogonal.
In the following we consider the solutions of \eq{Psieqgeneral} 
which are `periodic' and smooth in $\vect{\phi}$.

\vskip 2ex
  Now, let us list a few lemmas.

\lemma{a}{$\abs{\lambda}=1$}

\lemma{b}{$(\Psi_1,\Psi_2)=0$ if $\lambda_1\neq\lambda_2$}
  From the unitarity of $M(\vect{\phi})$ we get
\bdis
   \Psi_i^{\dagger}(\vect{\phi}) \Psi_j(\vect{\phi})
    &=& \Psi_i^{\dagger}(\vect{\phi}) 
     M(\vect{\phi})^{\dagger}M(\vect{\phi}) \Psi_j(\vect{\phi})
    =  
    \left[ M(\vect{\phi})\Psi_i(\vect{\phi}) \right]^{\dagger}
     M(\vect{\phi}) \Psi_j(\vect{\phi})   \\
    &=& \lambda_i^{*}\lambda_j \Psi_i^{\dagger}(\vect{\phi}+\vect{\mu}) 
             \Psi_j(\vect{\phi}+\vect{\mu}).
\edis
Integrating over $\vect{\phi}$ and using the definition
(\ref{defscalarprod}), we get [a] for $i=j$ 
and [b] for $\lambda_i\neq \lambda_j$. Note that [b] does not imply 
$\Psi_1(\vect{\phi})^\dagger \Psi_2(\vect{\phi})=0$ 
for $\lambda_i\neq \lambda_j$.

\lemma{c}{$\abs{\Psi(\vect{\phi})}$ is independent of $\vect{\phi}$
(and can be normalized to unity).}
The unitarity condition $\abs{\Psi(\vect{\phi})}=\abs{\Psi(\vect{\phi}+\vect{\mu})}$, 
together with the smoothness
of $\Psi(\vect{\phi})$, and the non-commensurability of $\vect{\mu}$
are enough to guarantee [c].

\lemma{d}{If $(\lambda,\Psi(\vect{\phi}))$ is a solution,
   so is $(\lambda^{*},\widehat{\Psi}(\vect{\phi}))$}
Take the complex conjugate of \eq{Psieqgeneral}
and use $\sigma_2 M^{*} \sigma_2 = M$. If $\Psi$ corresponds to
$\vect{n}$, then $\widehat{\Psi}$ corresponds to $-\vect{n}$ and the
spin tune changes sign. 
(Since $\sigma_2\vect{\sigma}\sigma_2=-\vect{\sigma}^{*}$,
${\widehat{\Psi}}^{\dagger}\vect{\sigma}\widehat{\Psi}
=-\Psi^{\dagger}\vect{\sigma}\Psi$.)
Note that not only $(\widehat{\Psi},\Psi)=0$ but also
$\widehat{\Psi}^\dagger\Psi=0$ at every $\vect{\phi}$.

\lemma{e}{If $\lambda$ is an eigenvalue, then so is
$\lambda e^{i\vect{m}\scp\vect{\mu}/2}$, where $\vect{m}$ is a set of
any integers.}
Multiply \eq{Psieqgeneral} by $e^{-i\vect{m}\scp\vect{\phi}/2}$ and define
$\widetilde{\Psi}(\vect{\phi})\equiv e^{-i\vect{m}\scp\vect{\phi}/2}
\Psi(\vect{\phi})$.
Then
\bdis
      M(\vect{\phi})\widetilde{\Psi}(\vect{\phi})=\lambda
         e^{-i\vect{m}\scp\vect{\phi}/2}\Psi(\vect{\phi}+\vect{\mu})
     =\lambda e^{i\vect{m}\scp\vect{\mu}/2}
      \widetilde{\Psi}(\vect{\phi}+\vect{\mu})
\edis
Thus, $\widetilde{\Psi}$ is an eigenvector belonging to the eigenvalue
$\lambda e^{i\vect{m}\scp\vect{\mu}/2}$.

   This gives an ambiguity in the spin tune: 
$\mu_s\rightarrow\mu_s+\vect{m}\scp\vect{\mu}$.
However, all the eigenvalues
of the form $\lambda e^{i\vect{m}\scp\vect{\mu}/2}$ give the same vector
$\vect{n}=\widetilde{\Psi}^{\dagger}\vect{\sigma}\widetilde{\Psi}
=\Psi^{\dagger}\vect{\sigma}\Psi$.
The $\vect{u}_{1,2}$ axes corresponding to $\widetilde{\Psi}$ are
\bdis
  \widetilde{\Psi}_{\varphi}=\frac{1}{\sqrt{2}} \left[
     e^{-i\varphi/2}e^{-i\vect{m}\scp\vect{\phi}/2} \Psi
     +e^{i\varphi/2} e^{i\vect{m}\scp\vect{\phi}/2} i\sigma_2 \Psi^{*}
     \right] = \Psi_{\varphi + \vect{m}\scp\vect{\phi}}
\edis
Thus, the new $\vect{u}_{1,2}$ axes rotate by $\vect{m}\scp\vect{\phi}$
with respect to the original ones.

\vskip 1ex
   From the lemmas above, we know that once a solution $(\lambda,\Psi)$
is found, we can construct infinitely many solutions of the form
$(\lambda e^{i\vect{m}\scp\vect{\mu}/2}, 
e^{-i\vect{m}\scp\vect{\phi}/2}\Psi)$ and 
$(\lambda^{*} e^{-i\vect{m}\scp\vect{\mu}/2}, 
e^{i\vect{m}\scp\vect{\phi}/2}\widehat{\Psi})$, and that
they all correspond to the same vector $\vect{n}$ or $-\vect{n}$.

   A natural question is then `are there any other eigenvalues?'. 
The answer is `No':
\lemma{f}{ If $\lambda$ is an eigenvalue, all other eigenvalues
are either  $\lambda e^{i\vect{m}\scp\vect{\mu}/2}$ or
$\lambda^{*} e^{-i\vect{m}\scp\vect{\mu}/2}$
}
If $(\lambda_1,\Psi_1)$ and $(\lambda_2,\Psi_2)$ are solutions,
$a(\vect{\phi})\equiv \Psi_2^\dagger(\vect{\phi})\Psi_1(\vect{\phi})
=(M\Psi_2)^\dagger M\Psi_1 = \lambda_2^{*}\lambda_1
a(\vect{\phi}+\vect{\mu})$. From the periodicity and smoothness of $a(\vect{\phi})$
and the non-commensurability of $\vect{\mu}$ one finds either that
[1] $a(\vect{\phi})=e^{i\vect{\alpha}\scp\vect{\phi}}$
and $\lambda_2^{*}\lambda_1=e^{-i\vect{\alpha}\scp\vect{\mu}}$,
$\vect{\alpha}$ being a constant 3-vector, or that
[2] $a=0$ ($\Psi_1$ and $\Psi_2$ are locally orthogonal).
In the case [1] $\vect{\alpha}$ must be of the form $\vect{m}/2$ from the
periodicity requirement, where $\vect{m}$ is a set of three integers.
Therefore, $\lambda_2=\lambda_1e^{i\vect{m}\scp\vect{\mu}/2}$.
In the case [2], examine
$\widehat{\Psi}_2$ in place of $\Psi_2$. Then we get either
$\lambda_2=\lambda_1^{*}e^{-i\vect{m}\scp\vect{\mu}/2}$
or $\widehat{\Psi}_2^\dagger\Psi_1=0$.
However, if both $\Psi_2^\dagger\Psi_1$ and 
$\widehat{\Psi}_2^\dagger\Psi_1$ vanish, then $\Psi_1=0$ because
$\Psi_2$ and $\widehat{\Psi}_2$ are orthogonal. Therefore
$\widehat{\Psi}_2^\dagger\Psi_1=0$ cannot be the case.
Thus, the cases [1] and [2] correspond to 
$\lambda_2 = \lambda_1 e^{i\vect{m}\scp\vect{\mu}/2}$ and
$\lambda_1^{*} e^{-i\vect{m}\scp\vect{\mu}/2}$, respectively.

\vskip 1ex
  Let us consider the spin tune $\nu_s=\mu_s/2\pi$. 
It is obtained from the definition
$\lambda=e^{-i\mu_s/2}=e^{-i\pi \nu_s}$. From the above arguments we
find that if the set $[\nu_s, \vect{n}, \vect{u}_1+i\vect{u}_2]$ 
is a solution, then
$[\nu_s-\vect{m}\scp\vect{\nu}, \vect{n},
 e^{-i\vect{m}\scp\vect{\phi}}(\vect{u}_1+i\vect{u}_2)]$
and $[-\nu_s-\vect{m}\scp\vect{\nu}, -\vect{n},
 -e^{i\vect{m}\scp\vect{\phi}}(\vect{u}_1+i\vect{u}_2)]$ 
are also solutions. Thus, the spin tune has ambiguities up to a
multiple of the orbit tunes and up to a sign. The latter is
related to the choice of sign of $\vect{n}$.

   When obtaining $\nu_s$ from $\lambda$, one finds an ambiguity
only up to an even integer rather than up to an integer. 
At first sight this is puzzling but it is also
due to the `2-to-1' correspondence between SU2 and SO3. Obviously,

\lemma{g}{If $\Psi$ is an eigenvector of $M$ with eigenvalue $\lambda$,
 it is also an eigenvector of $-M$ with eigenvalue $-\lambda$.}
 Since $M$ and $-M$ represent the same SO3 rotation,
 we have also to include the solutions to $-M$. However, $-M$ has
exactly the same eigenvectors as $+M$ (therefore the same 
$\vect{u}_1,\vect{u}_2,\vect{n}$) with spin tunes $\nu_s$ shifted
by one. This solves the above puzzle.  
Thus, we can define the spin tune
in the interval [0,1) or (-0.5,0.5] and, if the sign of $\vect{n}$
is irrelevant, we can reduce the interval to [0,0.5].

\vskip 1ex
  Thus, we have found that we only need one of the sets of eigenvector 
and eigenvalue of \eq{Psieq}.
All others can be constructed from this.
Therefore one can Fourier expand $\Psi$ as in \eq{FourierEq}
rather than as in the general from (\ref{shiftedFourier}). (Note, however,
that one will find tunes of the form $\pm\nu_s+2\vect{m}\scp\vect{\nu}$
although odd-multiple solutions can be easily reconstructed.)

\vskip 1ex
Let us briefly discuss the degeneracy. Within the eigenvalue 
group $\lambda e^{i\vect{m}\scp\vect{\mu}/2}$ of the
same sign of $\vect{n}$, a degeneracy is possible when
$\vect{m}\scp\vect{\nu}$ is an even integer, i.e., when the orbit motion
is in resonance, which we are not interested in. We may assume this is not the case.

   On the other hand, a degeneracy between solutions of different
signs of $\vect{n}$ corresponds to a spin-orbit resonance. When the two solutions
$(\lambda,\Psi)$ and $(\lambda^*,\widehat{\Psi})$ degenerate 
($\lambda=\lambda^*$),
the spin tune becomes an integer. Taking into account the ambiguity
of spin tune, this is equivalent to the relation  
$\nu_s=\vect{m}\scp\vect{\nu}+\mbox{integer}$.

\section{Choice of the Spin Tune}
  We have shown in the previous section that there are many eigenvalues
(spin tunes) representing the same vector $\vect{n}$ and different
$(\vect{u}_1,\vect{u}_2)$ axes.
Now, we must finally decide which eigenvalue to choose for the spin tune.
Theoretically speaking there is no reason to choose one particular
value. As pointed out in \cite{KYred} spin tune is intrinsically 
ambiguous up to a multiple of the orbit tunes.
The choice of the spin tune, which is equivalent to a choice of
$(\vect{u}_1,\vect{u}_2)$, is a matter of convention. 

In practice, however, a solution is not desirable if 
$(\vect{u}_1,\vect{u}_2)$ is a strong function of $\vect{\phi}$.
When one solves the equation by
Fourier expansion, the most natural choice is to take
the solution having the largest zero-Fourier harmonic
$\abs{\Psi_{\vect{m}=(0,0,0)}}$.

  If one plots all the eigenvalues as a function of any parameter 
(beam energy, betatron amplitude, etc), one will find continuous
curves. If one plots the spin tune selected as just described
as a function of these parameters
one may occasionally find a jump of spin tune although the whole spectrum 
content is continuous.

  Let us give an example from a test calculation. The test ring consists of 
100 FODO cells, each of which has two thin-lens quadrupole magnets
and two bending magnets filling the entire space between quadrupoles.
The focusing effect of the bending magnets is ignored.
 In order to avoid a too high symmetry of the orbit
motion, an artificial phase advance of 90 degrees in both horizontal and
vertical planes is introduced at one point in the ring.
The tunes are $\nu_x=15.3827$ and $\nu_y=25.6482$. 
Only the vertical betatron oscillation is excited.
The beam energy is so chosen that $\nu_{s0} = \gamma a=1520.72$.

Fig.1 shows the eigenvalue (spin tune) spectrum as a function
of the betatron action $J_y$. Only those with small $\vect{m}$ are plotted.
The points linked by a solid line
correspond to the spin tune selected by the criterion mentioned above.
As one can see, each eigenvalue is a continuous function of
$J_y$ (A few curves appear broken because not all the eigenvalues are plotted.)
but the selected tune shows a jump at 
$J_y\approx 0.7\times 10^{-8}$m$\cdot$rad. The spin tune
before and after the jump, $\nu_{s1}$ and $\nu_{s2}$ satisfies
$\nu_{s1}+\nu_{s2}=-2\nu_y+\mbox{integer}$.
The dashed line (with the same scale) is the upper limit of polarization, i.e.,
\beq
  P_{lim}=\abs{\avr{\vect{n}}}, \qquad
  \avr{\vect{n}} = \frac{1}{2\pi} \int_{0}^{2\pi} \vect{n}(\phi_y) d\phi_y
\eeq
 The minimum of $P_{lim}$
coincides with the point of the spin tune jump.
\begin{figure}[htb]
\epsfig{figure=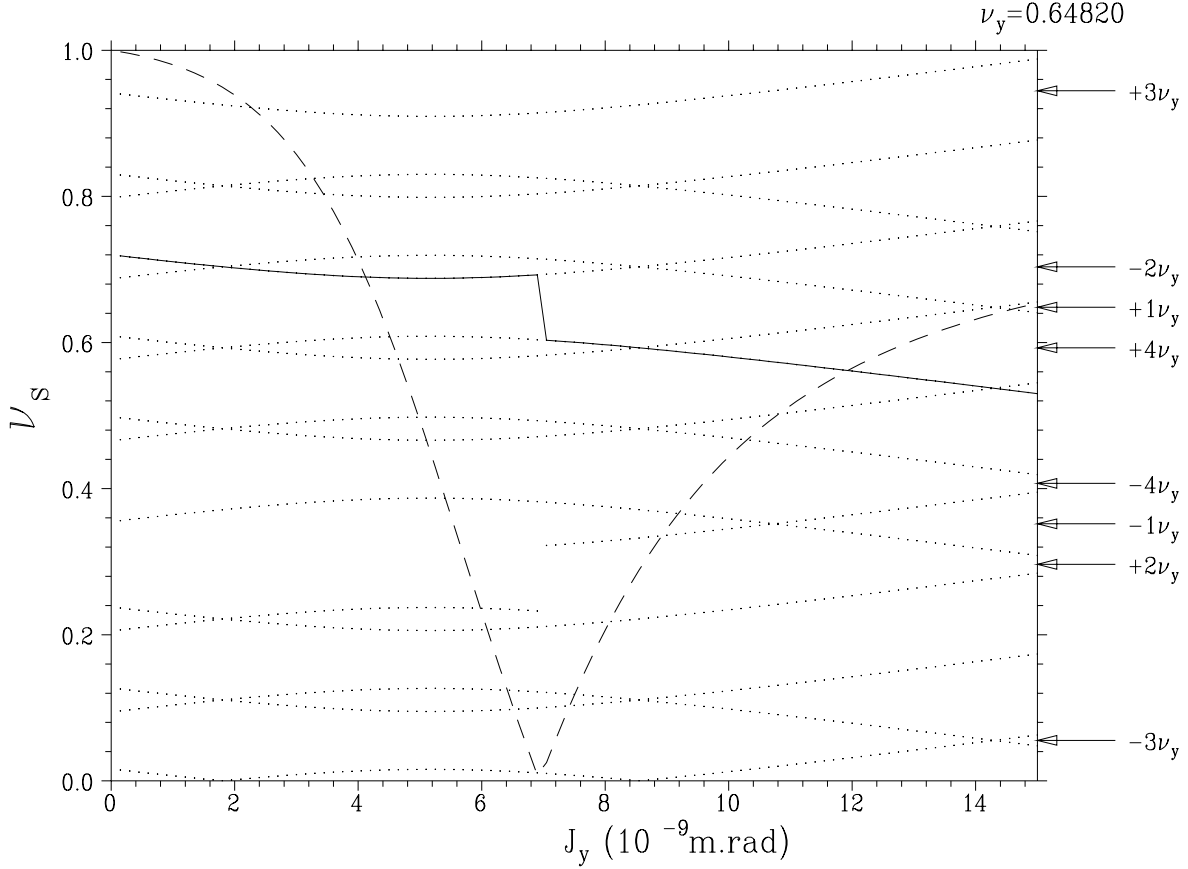,width=16cm}
\caption{An example of a spin tune spectrum as a function of betatron
action. The dashed line is the upper limit of polarization $P_{lim}$.}
\end{figure}

   We have compared the results of our program with SPRINT for the
amplitude dependence of the spin tune in the HERA ring. The agreement of 
the $\vect{n}$ axis and the spin tune was excellent.
Not only the occurrence of spin tune jumps but also their location agree, 
which means that taking the largest zero-harmonic
and the stroboscopic averaging are almost equivalent.

\section{Truncation of Fourier Series}
   In numerical calculations one has to truncate the Fourier expansion. 
There are a few problems associated with the truncation.

When $N$ values of $\phi$ are used (we deal with one degree of freedom
for illustration. The extension to 3 degrees of freedom is obvious.), 
the range of the harmonics should be $-\M\leq m\leq \M$ 
($N=2\M+1$).\footnote{
If $N$ is even, we have to change the upper or lower limit by 
one.}
  For a discrete Fourier transform the range
can also be $0\leq m\leq N-1$ (as in standard FFT routines), 
but this choice is not good when 
other values of $\phi$ are needed (for example when calculating
$\vect{n}$ for arbitary values after the problem is solved).
$N$ must be large enough to ensure that
the Fourier components  $M_m$ (actually $f_m$ and 
$g_m$) are small enough outside the region $[-\M,\M]$.\footnote{
This is not a sufficient condition for accuracy. Even if $M_m$ is small 
outside $[-\M,\M]$, the solution $\Psi_m$ can still be large in some cases.}

  The matrix $e^{-im\mu} M_{m-m'}$ 
in \eq{mateq} is then a $(2\M+1)\times(2\M+1)$ matrix (each element
is a $2\times2$ matrix and we are dealing with one degree of freedom.). 
One finds that the diagonal elements ($m=m'$) are 
normally large and that the elements with large $\abs{m-m'}$ are small.
The elements in the upper-right and lower-left triangles ($\abs{m-m'}>\M$)
are  exactly zero because they require the harmonics outside $[-\M,\M]$.
Owing to this truncation, the matrix does not exactly satisfy the lemmas
in the previous section even if $N$ is very large. (For example, in
the first row ($m=-\M$) even the first harmonic $m-m'=1$ is lost
because $m'$ would be $-\M-1$.) 
  Although the solution with the largest
zero-harmonic is not affected much by this truncation, it is not easy
to confirm the accuracy of the solutions.

   On the other hand,
 one can fill the upper-right and lower-left triangles 
by treating the harmonics in a
cyclic manner as in a discrete Fourier transformation (i.e.~one
identifies the ($\M+1$)-th harmonic with the ($-\M$)-th.).  With this
prescription the truncated matrix becomes exactly unitary even if $N$ is not 
large enough.
The solution with the largest zero-harmonic does not change much.
The appearance of eigenvalues with modulus far from unity then means that 
the eigenvalue solver is not accurate.

   When one adopts the cyclic use of the harmonics, the lemmas
[a], [b], [d] and [g] hold exactly apart from round off errors,
but [c] and [e] (and accordingly [f]) become inaccurate.

\section{Conclusion}
We have shown that the spin tune can be obtained as an eigenvalue
of a matrix which is created from the one-turn maps
calculated by particle tracking. The method is applicable to any
system with linear or nonlinear orbit motion as long as the orbit action
variables exist. 
The convergence is
much better than with perturbation methods and the previous SODOM
algorithm. The computation is very 
fast because it  makes full use of the fact that the spin motion is
linear and that we know the orbit tunes.

{\bf Acknowledgements}
  The author thanks to Drs.~D.~Barber, K.~Heinemann, G.~Hoffstaetter,
and M.~Vogt for stimulating discussions.

\end{document}